\documentclass[10pt,a5paper]{article}
\usepackage[a5]{crop}

\usepackage{kopsavil}
\usepackage[cp1257]{inputenc}
\usepackage{graphicx}
\usepackage[T2A,OT1]{fontenc}
\renewcommand{\baselinestretch}{1}

\begin{document}
\hyphenpenalty = 10000
\crop
\large

\thispagestyle{empty}
{\centering

{\large \bf RIGA TECHNICAL UNIVERSITY}

{\large Faculty of Computer Sciences and Information Technologies

Institute of Information Technologies}

\vskip 1.5cm

{\large \bf Maxim FIOSHIN}

{\large Information Systems doctoral program doctoral student\\
Doctoral student card Nr. 001RDD003}

\vskip 15mm

{\large \renewcommand{\baselinestretch}{1.4}
\selectfont \bf RESAMPLING APPROACH FOR THE CALCULATION PROCESSES AND INFORMATION SYSTEMS MODELS ESTIMATION

}

\vskip 15mm

{\large \bf Promotion work summary}

\vskip 15mm

\begin{flushright}
{\large
Supervisor:\\
Dr.Habil.Sc.Eng., professor\\[0.2 cm]
A. ANDRONOV}
\end{flushright}

\vskip 30mm

{\large \bf Riga - 2005}

}

\newpage

\thispagestyle{empty}

\noindent
UDK 004 + 519.2](043)\\
Fi 748 r

\vskip 2cm

{\sloppy \parindent=35mm \hangindent=35mm \hbadness=10000
Fioshin M. Resampling Approach for the Calculation Processes and Information Systems
Models Estimation. Promotion work summary.-R.:RTU,2005.-34 pp.

}

\vskip 25mm

{\sloppy \parindent=35mm \hangindent=35mm \hbadness=2000
Printed according to an IT institute decision,
January 6, 2005, protocol Nr. 05-01.

}

\vskip 60mm

{\noindent \large \bf ISBN 9984-32-705-1}

\newpage

\begin{centering}
\bf
PROMOTION WORK
PRESENTED\\ TO THE RIGA TEHNICAL UNIVERSITY\\
TO OBTAIN THE SCIENTIFIC DEGREE OF\\
DOCTOR OF SCIENCE IN ENGINEERING\\
IN INFORMATION TECHNOLOGIES\\
\end{centering}

\vskip 3mm

\noindent
The promotion work presented to obtain the scientific degree of Doctor of Science in Engineering in Information Technologies is publicly defended April 11, 2005 at the
Riga Technical University Faculty of Computer Sciences and Information Technologies, Meza str.1, aud. 3-202.

\vskip 3mm

OFFICIAL OPPONENTS

\vskip 2mm

{\noindent
Dr.habil.sc.ing., professor Vyacheslav Melas\\
St. Petersburg State University}

\vskip 3mm

{\noindent
Dr.Sc.Eng., as. professor Irina Yatskiv\\
Transport and Telecommunications Institute}

\vskip 3mm

{\noindent
Dr.Habil.Sc.Eng., professor Juri Merkuryev\\
Riga Technical University}

\vskip 3mm

CONFIRMATION

\vskip 2mm

\noindent
I confirm, that I developed the present promotion work, that is presented
to the Riga Technical university to obtain the scientific degree of Doctor of Science in Engineering. The promotion work has not been presented to any other university to obtain a scientific degree.

\vskip 3mm

{\noindent
Maxim Fioshin .................................}

\vskip 2mm

{\noindent
Date: March 11, 2005}

\vskip 2mm

\noindent
The promotion work is written in English, contains an introduction, 6 sections, a conclusion,
bibliography, 27 tables, 30 figures and illustrations, 122 pages in total. 
The bibliography contains 97 entries.

\newpage

\begin{center}
\bf \Large
Abstract
\end{center}

\noindent
Resampling Approach for Estimation of Models of Calculation Processes and Information Systems.
M. Fioshin. The doctoral degree thesis. Supervisor Dr.Habil.Sc.Eng., professor A. Andronov. 

The work is devoted to the analysis of the Resampling method proposed by A. Andronov and to the analysis of the Resampling method application possibility to the estimation and simulation of the calculation and logical systems reliability. The work 
Simple and Hierarchical method properties are considered, algorithms for variance are shown. The methods are applied for processes in the multitask operation system and queries to database analysis, a comparison with the classical method, that uses the empirical distribution functions, is made. Numerical examples illustrate the influence of different factors on the Resampling method efficiency.

The task of the sample size optimization has been considered. The dynamic programming
method is applied to minimize the variance of the Resampling estimator. Optimization is applied for the analysis of queries to database, the numerical example illustrates 
the value of optimization.

The case of partially known distributions is considered. It is shown how to use
the Resampling approach in the case when the distributions of
some input variables are known. The method is applied to database query analysis and a
comparison with Hierarchical Resampling is made.

The construction of the Resampling confidence intervals is considered. The algorithm for
construction of the Resampling confidence intervals is shown and actual coverage
probabilities are calculated. Examples for the multitask operation system analysis illustrate 
the calculation of the actual coverage probability algorithm.

\newpage

\tableofcontents

\newpage

\section{Importance of the Work Subject}

At the present time the possibilities of computers are developing rapidly. The computer has become a common instrument of a scientist. It can help us in scientific research and allows us to solve tasks, that could not be solved before.

Thus, a question arises - how to use a computer in scientific research?
How can a computer help a scientist, besides simple calculations and information storing? A great attention is paid to this question at a present time.

When computers appeared, a new directions began to develop in many sciences 
which tried to solve the problems of respective science by using a computer.
At the beginning they were numerical methods in mathematics and physics. Later computers began to be used for problem solving in such sciences as chemistry, biology, geology, economics etc.

In early 70-s the possibilities of the computer started to be used also in statistics. It was clear, that by using computers data analysis can be efficiently performed. But classical statistical methods, as in the rest of sciences, are not oriented to computer application. Classical methods suppose formula is obtained as the method result, and the formula gives  result after small amount of calculations. Such method application is relatively complex, requires different assumptions about the model kind, requires model transformation, which is difficult to realize on a computer. 

As an alternative to classical methods a group of statistical methods appears, called intensive statistical computer methods or calculation statistics. The methods which belong to this group are simple, can be easily realized on the computer, but require a big amount of calculations. Usually these methods do not require many assumptions about the model structure, do not require complex data transformations, but the result is not so accurate, as in the case of classical methods.

As intensive methods do not require many assumptions about the model structure, they can be used to solve a wide class of problems. These methods allow us to analyze data from different points to discover dependences, that were not seen before. Intensive computer methods allow us to solve problems, which in the classical model limits cannot be solved or can be solved with big assumptions.

This area is rapidly developing. Now the amount of information is huge and analysis of information has become one of the most important tasks of computer sciences. In real situations we need to solve tasks which are difficult to solve using classical statistical methods. For  solving such tasks the intensive statistical computer methods are used.

The intensive statistical methods have two sides. On the one hand, the usage of such methods is simple. But, on the other hand, accurate analysis of such methods is a complex task. Often it is more difficult to analyze a simple intensive method than a complex classical one. But the analysis of intensive methods is required, because without it we cannot guarantee, that the method will give a correct answer and in the case of the correct answer there is no possibility to estimate the efficiency and accuracy of the answer. Thus the analysis of intensive statistical methods is actual task.

A new intensive statistical method, called Resampling, is considered in the work. This method can be applied efficiently for different statistical tasks solving, for example, statistical estimation, simulation, confidence intervals construction. The method can be applied for different systems estimation and simulation, including the analysis of information systems. Possible applications of the method for information systems estimation are considered in the work, application examples are shown.

One of the main goals of the work is efficiency analysis of the Resampling method  in the case when it is applied for the information systems simulation. This task is topical, because Resampling cannot be correctly applied without such analysis. 

\section{Goal and Tasks of the Work}

The goal of the work is obtaining algorithms for the calculation of Resampling method property efficiency, the application of the Resampling method for the information systems estimation and application of the algorithms for the method efficiency calculation.

The following issues are supposed to be the main tasks of the work:
\begin{itemize}
\item To study the Resampling approach and fields of its application.
\item Using simple and Hierarchical Resampling methodology, develop algorithms for the Resampling application for such tasks, as sample size optimization and the case of partially known distributions.
\item Develop algorithms for the method efficiency estimation in the mentioned cases.
\item Develop algorithms for the application of Resampling method for confidence intervals construction.
\item Develop algorithms that allow us to estimate the accuracy of Resampling confidence intervals.
\item Consider a possibility of applying the Resampling method in the information technology area.
\item Using the Resampling methodology, make estimations for different models from the information technology area and apply algorithms for the estimator efficiency calculation.
\end{itemize}

\section{Research Methodology}

As the  theoretical and methodical basis of the promotion work, the classical works in the computer science, simulation, statistics and probability theory were used.

Books in the corresponding areas, periodical publications of the
thematic materials, materials of international conferences in the corresponding areas were used in the promotion work .

During research, examples from information technology areas were analyzed,
in which concrete application of the developed methodology was illustrated. 
Hypothetical data was used in examples, which illustrate the specific character and efficiency of the method as fully as possible. As the method efficiency criterion the variance of estimator was used. The change of the method efficiency depending of different factors
was analyzed, which allows us to speak about the possibility of applying the method in practical situations.

For solving the given problems both analytical and experimental methods were applied.
Using the analytical methods analytical expressions for the method efficiency
calculation in different situations were obtained. Using experimental methods
the values of the methodic usage efficiency criterion for the concrete numerical
examples were calculated, which allow to see the different factors influence to the
method accuracy.

\section{Scientific Novelty of the Work}

Intensive statistical computer methods include many methods, such as the jackknife, Bootstrap and Resampling methods, and allow us to solve a wide class of problems. The Jackknife method was proposed by  Tukey in 1958 as an estimator which is a combination of an estimator based on all data and estimators based on parts of data. In 1979 Efron proposed the Bootstrap method, which in fact is generalization of jackknife. 

In 1976 Ivnitsky proposed to use Resampling for the tasks of reliability estimation. This approach has been developing since 1995 supervised by prof. Andronov. 
Andronov considered simple and Hierarchical Resampling methods,  Andronov, Merkuryev and Loginova considered application of the method for reliability and queuing theory, Andronov, Merkuryev and Fioshin considered Resampling method optimization tasks \cite{AndrF_99_2},
Andronov and Fioshin considered Resampling sum properties \cite{AndrF_98}, \cite{AndrF_99}, the case of partially known distributions \cite{AndrF_99_3}, confidence interval construction \cite{AndrF_04}, in the present time Andronov and Afanasjeva  work on the method application in regression analysis.

The application of the Resampling method for analysis of information systems has not been analyzed before. Different models from the information technology area are analyzed in the work (multitask operation systems, database queries, reliability of information storage), the methodology of different variants of Resampling method application for the considered models has been examined (simple Resampling, Hierarchical Resampling, Resampling in the case of partially known distributions), and also different tasks are considered (point estimation, interval estimation, sample size optimization).

Algorithms for Resampling method efficiency calculation for the considered models are constructed in the work. These algorithms can be applied for a wide class of problems and show how to estimate the efficiency of the method in similar situations.
In similar situations the efficiency of Resampling method can be analyzed using the same methodology. The results of the work can be used as the basis of the Resampling simulation software development.

\section{The Main Results of the Work}

The main results of the work are following:
\begin{itemize}
\item The methodology of the Resampling method application is considered for different cases  (Hierarchical Resampling, the case of partially known distributions) and for different tasks (the estimation of the expectation, optimization of sample sizes, confidence interval construction), which are described in articles of Andronov, Merkuryev and Fioshin;
\item Models from information technology area are selected and described, and the Resampling method can be used for their analysis;
\item It is shown, that the Resampling method can be applied for the estimation and simulation of such models;
\item Algorithms for the method efficiency criteria calculation are obtained for each concrete system;
\item Different method variants for concrete systems have been compared;
\item The influence of the system parameters on the method efficiency is analyzed and conclusions are drawn about the method application possibility for the given concrete system class.
\end{itemize}

\section{Practical Application of the Work}

Using the results obtained in the work it is possible to use the Resampling method for information system estimation. The methodology of the method application and efficiency calculation are shown in the work, which allows us to use Resampling in practical simulation. The obtained results make software construction possible, which makes Resampling estimation of different systems and correctly estimates the method efficiency, allowing correct experiment planning.

\section{Publications and Participation at Conferences}

The results of the work have been presented in 8 publications [1-8], and also presented in discussions at the corresponding conferences. 

\section{Structure of the Work}

In the first section of the work the intensive statistical computer methods are described, a short description is given. The Resampling method is also described and its application possibilities for the information system estimation are shown. In each of the following sections one case or task of the Resampling method application is considered. The 2-nd section describes the simple Resampling, the 3-d section describes the Hierarchical Resampling, the 4-th section describes the task of sample size optimization, the 5-th section describes the case of partially known distributions, the 6-th section describes the construction of Resampling confidence intervals. Tasks are described and algorithms are given. 
Then follows the method efficiency calculation. At the end of each section examples are considered. Next numerical results follow, which allow us to compare different method variants and analyze the influence of the system properties on the method efficiency criteria. At the end of each section conclusions are made about the efficiency of the method application for the given case or task.

\section{Short Description of the Work Sections}

\subsection{Intensive Statistical Computer Methods}

As the work is devoted to the Resampling method, which is one of the intensive statistical methods, in the first section of the work the intensive statistical methods analysis is performed.
At the present time some authors consider computational statistics a separate discipline.

It is often difficult to apply traditional statistical methods for complex systems modeling, non-stationary systems, cases when distributions differ from classical. In these cases it is better to use the intensive computational methods.

The intensive methods are simple, and it is simple to realize them. It is also simple to use such methods because few assumptions about model structure are required.

On the other hand, intensive methods do not give accurate results, as the classical methods do. The simplicity of these methods and the existence of many variants leads to many realizations and increase the possibility of incorrect method usage. One must remember that many computations not necessarily guarantee a correct result.

At the present time, 3 main intensive computer statistical methods are mentioned:

\begin{itemize}

\item The Monte Carlo methods.
\item Randomization methods, which include  cross-validation and the jackknife method.
\item Resampling methods.

\end{itemize}

A brief description of each method group is given in the section. 

Next a general description of the proposed Resampling method follows. 
The possible application spheres of the method in the information technologies area are shown.

The Resampling method can be used for the following problems in the information technology area:

\begin{itemize}
\item Database design and performance analysis.
\item Software reliability.
\item Server performance and efficiency analysis.
\item Multitask operating systems work optimization.
\item Network analysis and optimization.
\item Information protection.
\item Information storage device reliability analysis and information backup. 
\end{itemize}

The Resampling method can be successfully applied for system analysis, if the system has the following properties:
\begin{itemize}
\item A small amount of the input statistical information.
\item The analyzed events are relatively rare.
\item An unknown type of the system random value distributions.
\item A known functional dependence on initial data.
\end{itemize}

\subsection{Resampling Point Estimation of Calculation System Models}

Suppose we have independent random variables $X_1,X_2,\ldots,X_m$. 
The distribution functions $F_i(x)$ of these variables
are unknown, but the sample populations $H_i=\{X_{i1},X_{i2},\ldots,X_{in_i}\}$ 
are available for each variable $X_i$,  $i=1,\ldots,m$.

Suppose a known function $\phi(x_1,x_2,\ldots,x_m)$ of $m$ real arguments is given.
The task is to estimate the expectation $\theta$ of the function $\phi$, 
the arguments of which are random variables $X_1,X_2,\ldots,X_m$:

\begin{equation}
\theta = E_{\scriptscriptstyle F_1,F_2,\ldots,F_m}\phi(X_1,X_2,\ldots,X_m).
\end{equation}

The traditional estimation methods usually propose the so-called "plug-in" procedure.
It means that instead of the real distribution function $F_i(x)$ its estimators 
$\widehat{F_i}(x)$ are used (as the estimators the empirical distribution functions
are often used). Then the estimator $\widehat\theta$ of $\theta$ is following:

\begin{equation}
\widehat\theta = E_{\scriptscriptstyle \widehat F_1,\widehat F_2,\ldots,\widehat F_m}\phi(X_1,X_2,\ldots,X_m).
\end{equation}

The idea of the method application is following. We select at random an element from each sample $H_i$. Suppose at the step number $l$ the element with number $j_i(l)$ is extracted from sample $H_i$. Let us create a vector $X(l)$ from elements extracted on $l$-th step: 
$X(l)=(X_{1j_1(l)},X_{2j_2(l)},\ldots,X_{mj_m(l)})$.

Let us repeat this procedure $r$ times, obtaining realizations $X(1), X(2), \ldots, X(r)$.
The estimator $\theta^*$ of the value $\theta$ is equal to an average of the function $\phi$ on all $r$ realizations:
\begin{equation}
\theta^* = \frac{1}{r}\sum_{l=1}^{r}\phi(X(l)).
\end{equation}

It is proved, that the estimator $\theta^*$ is unbiased: $E\;\theta^*=\theta$.

Let us take the estimator $\theta^*$ variance as the method efficiency criterion.
Let $\mu = E\;\phi(X(l))$; $\mu_2 = E\;\phi(X(l))^2$; $\mu_{11} = E\;\phi(X(l))\phi(X(l'))$,
where $l$ and $l'$ are realization numbers. Using properties of variance, we have:
\begin{equation}
D\;\theta^*=\frac{1}{r}(\mu_2+(r-1)\mu_{11})-\mu^2.
\end{equation}

Only the mixed moment $\mu_{11}$ depends on the element extraction rules.

In order to calculate $\mu_{11}$, we use the $\omega$-pair notation. We will say, that vectors
$j(l)$ and $j(l')$ produce the $\omega$-pair, if $j_i(l)=j_i(l')\Leftrightarrow i\in\omega$,
or, in other words, the set $\omega$ contains numbers of elements, which are equal in samples $X(l)$ and $X(l')$. For example, vectors $(2,1,4,2)$ and $(2,2,4,1)$ produce the $\{1,3\}$-pair.

Let us suppose a $\mu_{11}(\omega)$ is a conditional mixed moment by the condition, that the $\omega$-pair takes place.
Let us suppose $P\{\omega\}$ is the probability to get the $\omega$-pair. Then the value of $\mu_{11}$ can be calculated as following:
\begin{equation}
\mu_{11} = \sum_{\omega}P\{\omega\}\mu_{11}(\omega).
\end{equation}

\underline{Example 1: The reaction time of an information system.}
Let us have a calculation system, the reaction time of which depends on some parameter $X$ 
($X$ can be the size of input data in an algorithm,
the size of the database for a database management system, the number of processes in
a computer when the next process is created etc.). We suppose that $X$ is a random variable,
its distribution $F(x)$ is unknown, but 
the sample $H=(X_1,X_2,\ldots,X_n)$ of $X$ realizations is available.

In this case the function $\phi$ depends on one argument $x$. The task is to estimate the expectation of this function. 
\begin{equation}
\theta = E\;\phi(X).
\end{equation}

We can use the Resampling method in order to estimate $\theta$.
The variance of estimator $\theta^*$ is calculated. The comparison results are shown
in table \ref{t2_1_1}. We can see that the variance of Resampling estimator is 10-15\%
greater than the classical estimator variance, but the application of Resampling is simpler,
than the application of classical methods.

\begin{table}[ht]
\caption{Variance dependence on sample size $n$}
\label{t2_1_1}
\centering
\begin{tabular}{|c|c|c|}
\hline
\begin{tabular}{c}
Sample size, \\
$n$\\
\end{tabular}& 
\begin{tabular}{c}
Classical estimator\\ variance, $D\;\hat{\theta}$
\end{tabular}& 
\begin{tabular}{c}
Resampling estimator\\ variance, $D\;\theta^*$
\end{tabular}\\
\hline
1& 781.25& 781.25\\
\hline
2& 390.625& 398.437\\ 
\hline
3& 260.417& 270.833\\ 
\hline
5& 156.25& 168.75\\
\hline
8& 97.6562& 111.328\\ 
\hline
10& 78.125& 92.1875\\ 
\hline
13& 60.0962& 74.5192\\ 
\hline
15& 52.0833& 66.6667\\
\hline
\end{tabular}
\end{table}

\underline{Example 2: sequential processes.} 
Let us assume the task consists of $m$ sequential processes.
The random variables $X_1, X_2, \ldots, X_m$ are the process execution times. The distributions $F_i(x)$ of these variables are unknown, but only samples $H_i$ of each process execution time are available. We need to estimate the average time of the task execution.

In this case the function $\phi$ is the sum of variables $X_i$. We need to estimate the expectation of this sum:
\begin{equation}
\theta = E\;\ X_1+X_2+\ldots+X_m.
\end{equation}

Formulas for $\mu_{11}$ value calculation are obtained. Variance dependence on different parameters is analyzed, different cases are compared and it is shown that the method is
relatively effective for solving this task. Variance dependence on sample sizes $n$ is shown on the  Fig. \ref{tree3}.

\fig{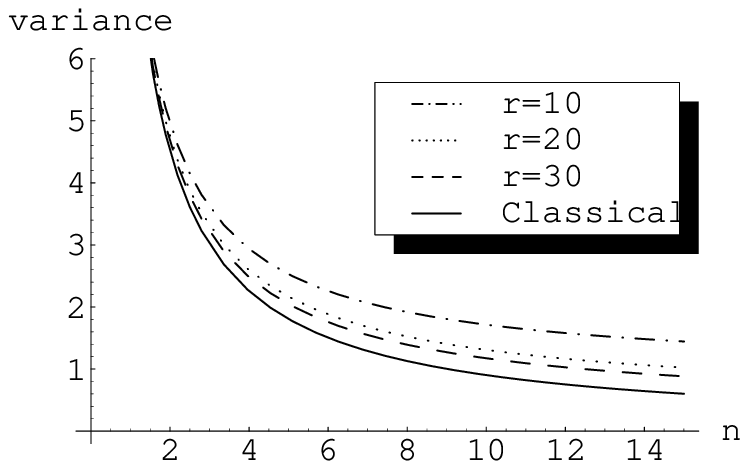}{Variance dependence on sample size $n$}{tree3}

\underline{Example 3: Parallel processes.} 
Suppose the task consists of $m$ parallel processes.
The random variables $X_1, X_2, \ldots, X_m$ are the process execution times. The distributions $F_i(x)$ of these variables are unknown, but only samples $H_i$ of each process execution time are available. We need to estimate the average time of the task execution.

In this case the function $\phi$ is maximum of variables $X_i$. We need to estimate the expectation of this function:
\begin{equation}
\theta = E\;\max(X_1,X_2,\ldots,X_m).
\end{equation}

Formulas for $\mu_{11}$ value calculation are obtained. Variance dependence on different parameters is analyzed, different cases are compared and it is shown that the method is
relatively effective for this task. 

\begin{table}[ht]
\caption{Variance dependence on sample sizes $n$}
\label{t2_2_1}
\centering
\begin{tabular}{|c|c|c|c|c|}
\hline
\begin{tabular}{c}
Sample\\ size,\\
$n$ 
\end{tabular}& 
\begin{tabular}{c}
Resamples\\ count,\\ r=10
\end{tabular}&
\begin{tabular}{c}
Resamples\\ count,\\ r=20
\end{tabular}&
\begin{tabular}{c}
Resamples\\ count,\\ r=30
\end{tabular}&
\begin{tabular}{c}
Variance of\\ classical\\ estimator,\\ $D\;\hat{\theta}$
\end{tabular}\\ 
\hline
1& 9.02778& 9.02778& 9.02778& 9.02778\\
\hline
2& 4.96528& 4.73958& 4.66435& 4.51389\\ 
\hline
3& 3.61111& 3.31019& 3.20988& 3.00926\\ 
\hline
5& 2.52778& 2.16667& 2.0463& 1.80556\\
\hline
8& 1.9184& 1.52344& 1.39178& 1.12847\\ 
\hline
10& 1.71528& 1.30903& 1.17361& 0.902778\\
\hline
12& 1.57986& 1.16609& 1.02816&  0.752315\\
\hline
15& 1.44444& 1.02315& 0.882716& 0.601852\\
\hline
\end{tabular}
\end{table}

\underline{Example 4: Reliability of information storage} 
Suppose the information storage consists of 3 reservation devices.
We say that the system is reliable if at least 2 of 3 devices work. 
The working times of the devices before failure are independent 
random variables $X_1$, $X_2$ and $X_3$. The distributions
$F_1(x)$, $F_2(x)$ and $F_3(x)$ of the device working time are unknown,
only sample populations $H_1$, $H_2$ and $H_3$ are available.
The task is to estimate the probability that at the time moment
$t$ the system is reliable.

In this case the function $\phi_t(x_1,x_2,x_3)$ is the indicator function  
which returns to 1 if the system works at time $t$ and to 0 if the systems fails,
if working times of elements are $x_1,x_2,x_3$ correspondently. 
The function $\phi_t$ can be defined as follows:
\begin{equation}
\phi_t(x_1,x_2,x_3)=\left\{\begin{array}{rl}
1 &\mbox{if at least 2 el-ts of $\{x_1,x_2,x_3\}$ are $>t$},\\
0 &\mbox{otherwise. }
\end{array}\right.
\end{equation}
\par
The goal is to estimate the expectation $\theta_t$ of this function:
\begin{equation}
\theta_t = E\;\phi_t(x_1,x_2,x_3).
\end{equation}
It is clear that $\theta_t$ is the probability that at the time moment $t$ the system is reliable.

Formulas for $\mu_{11}$ value calculation are obtained. Variance dependence on different parameters is analyzed, different cases are compared and it is shown that the method is
relatively effective for this task. Variance dependence on time $t$ is shown on Fig. \ref{tree7}.

\fig{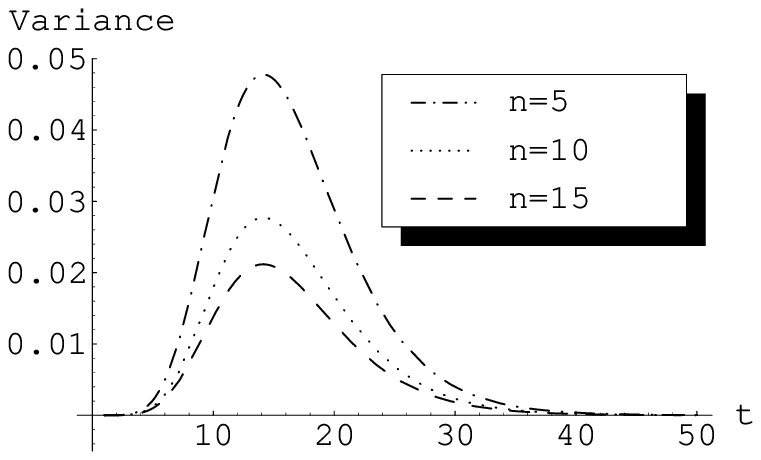}{Variance dependence on time $t$}{tree7}

\subsection{Hierarchical Resampling for the
Point Estimation of Hierarchical Calculation Systems}

Hierarchical Resampling has the following advantages in comparison with the simple Resampling:

\begin{itemize}
\item The method allows to accomplish simpler estimation of complex systems, which consist of subsystems.
\item The method allows us to perform parallel calculations for the subsystems analysis.
\item The method allows to accomplish optimization of sample sizes.
\item The method can be applied for complex information systems analysis, such as hierarchical queues to databases, enterprise databases, hierarchical servers structures etc.
\end{itemize}

Suppose function $\phi(x_1,x_2,\ldots,x_m)$ can be represented by using subfunctions $\phi_j(\cdot)$.
The result of the subfunction is used as the value of higher level function argument.
In this case the function $\phi(x_1,x_2,\ldots,x_m)$ can be represented by using the calculation tree.

\begin{figure}[htbp]
	\centering
		\includegraphics[width=0.50\textwidth]{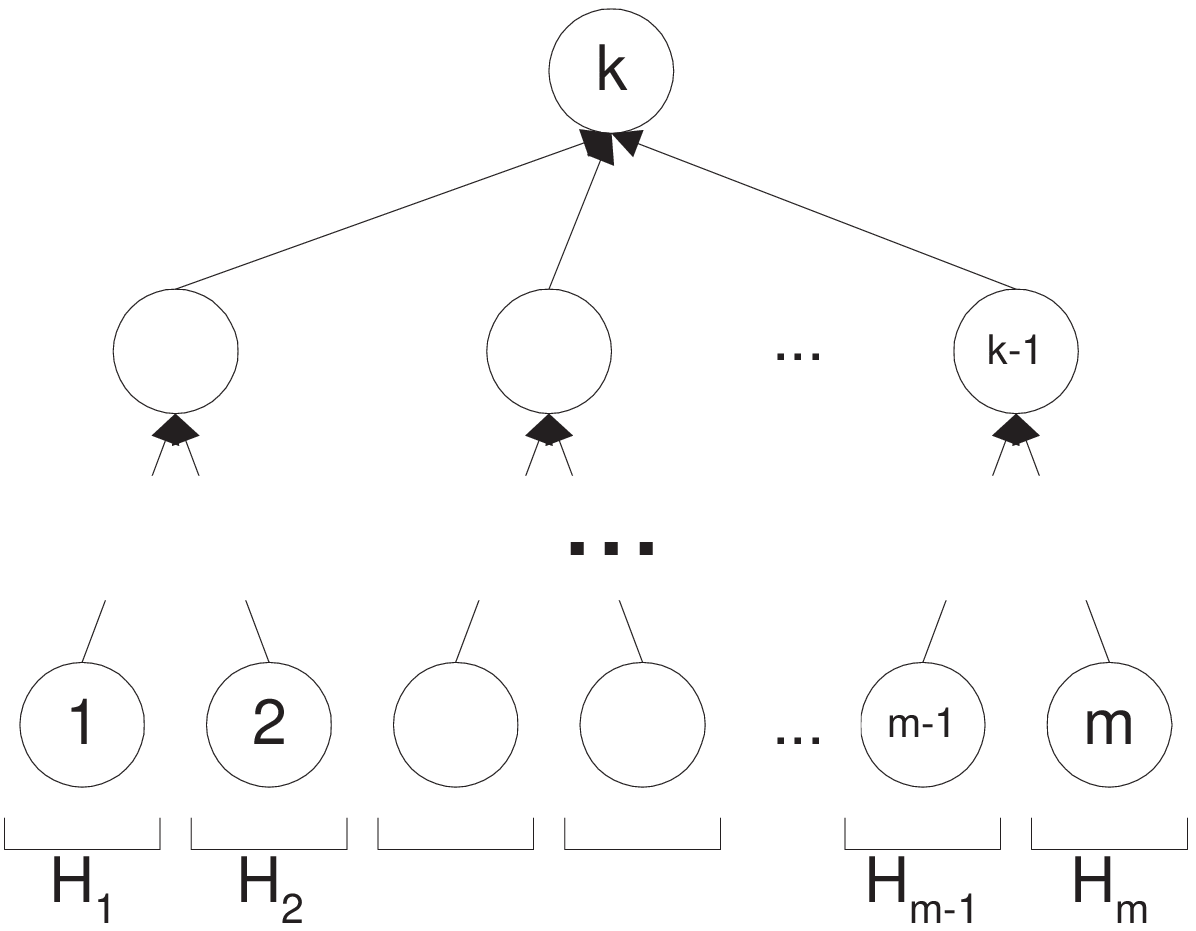}
	\caption{Calculation tree}
	\label{fig:calctree}
\end{figure}

The input variables $X_1,X_2,\ldots,X_m$ correspond to the tree leaves. The rest of the nodes are intermediate ones, and intermediate functions $\phi_j(\cdot)$ correspond to them. The result of each function is taken as an argument of the function on a higher level. The function $\phi(x_1,x_2,\ldots,x_m)$ correspond to the root of the tree. An example of the calculation tree is presented on Fig. \ref{fig:calctree}.

A sample $H_v$ corresponds to each node $v$. During the simulation the samples are constructed iteratively, by levels. The total estimator $\theta^*$ of the value $\theta$ is
equal to the average value at the root of the tree:
\begin{equation}
\theta^* = \frac{1}{n_k}\sum_{l=1}^{n_k} Y_{kl},
\end{equation}
where $Y_{kl}$ are elements of the sample $H_k$.

Let us take variance $D\;\theta^*$ of the estimator $\theta^*$ as the method efficiency criterion.
The variance calculation is based on the $\omega$-pair definition.
The probabilities of $\omega$-pairs and conditional mixed moments $\mu_{11}(\omega)$ are calculated iteratively, by the tree levels.

\underline{Example 1: Hierarchical query to database.}
Let we have a query to database that consists of 6 subqueries. A subquery $i$
working time is a random variable $X_i$, $i=1,\ldots,6$. The distributions
$F_i(x)$ of subquery working times are unknown, only sample populations
$H_i$ are available for each $i$. 

The query is executed on 3 processors (or database servers). The execution rules
are the following:
\begin{itemize}
\item 1-st and 2-nd subqueries are executed on the 1-st processor, and they are executed in parallel;
\item 3-d and 4-th subqueries are executed on the 2-nd processor, and they
are also executed in parallel, but the 2-nd processor ends its work when one of
the subqueries gives a result;
\item 5-th and 6-th subqueries are executed on the 3-d processor, and they are executed sequentially.
\end{itemize}

The task is to estimate probability $\theta_t$, that to the time moment $t$ the query will end its work, which can be written as following:
\begin{equation}
\label{eq12}
\begin{array}{c}
\theta_t = P\{\max(\max(x_1,x_2),\min(x_3,x_4),x_5+x_6)<t\} = \\
=E\;\phi_t(x_1,\ldots,x_6).
\end{array}
\end{equation}

Formulas for $\mu_{11}$ value calculation are obtained. Variance dependence on different parameters is analyzed, different cases are compared and it is shown that the method is
relatively effective for this task. Variance dependence on time $t$ is shown on Fig. \ref{draw3_4}.

\fig{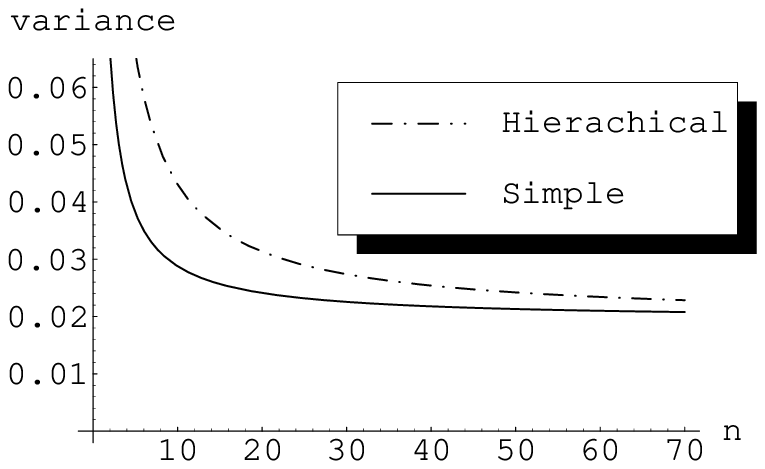}{Comparison of simple and Hierarchical Resampling methods}{draw3_4}

\underline{Example 2: Sequential - parallel query to database}

Let we have a query to database, the subqueries of which are organized in blocks. 
All subqueries in one block
are executed in parallel. The block gives the result when all subqueries
in the block give the result. The query gives the result when the
first block gives the result.

Let us assume all subqueries inside a block is the same distribution of the working time.
The distribution function $F_i(x)$ of the subqueries working time is unknown, but only sample 
$H_i$ is available. Only one sample is available for each block.

The goal is to estimate the probability $R(t)$, that the query working time is greater than  $t$: $R(t)=P\{X>t\}$, where $X$ is the working time of the query. 
If the distribution of the subqueries working time is known, then $R(t)$ can be calculated in the following way:
\begin{equation}
\label{ex25_1}
R(t)=\prod_{i=1}^n \left(1-F_i(t)^{l_i}\right),
\end{equation}
where $F_i(t)$ - is the distribution function of the block $i$ subquery working time.

If we use the empirical distribution function $\hat{F_i}(t)$ for the $R(t)$ value estimation, we get the following estimator:
\begin{equation}
\label{ex25_3}
\hat{R}(t) = \prod_{i=1}^n (1-\hat{F_i}(t)^{l_i}).
\end{equation}

It is shown that the estimator (\ref{ex25_3}) is biased. The dependence of bias on the time 
$t$ is shown in table \ref{t1_25}.

\begin{table}[ht]
\centering
\caption{The expectation and bias (\%) of the traditional estimator $\hat{R}^*(t)$ depending on time $t$}
\begin{tabular}{|l|l|l|l|l|l|}
\hline
$t$ & 0.1	& 0.2	& 0.3	& 0.5	& 0.7\\
\hline
$R(t)$	& 0.999	& 0.987	& 0.954	& 0.818	& 0.629\\
\hline
$E\;\hat{R}^*(t)$	& 0.992	& 0.960	& 0.901	& 0.723	& 0.523\\
\hline
\%	& 1\%	& 3\%	& 6\%	& 13\%	& 20\%\\
\hline
\hline
$t$	& 0.9	& 1	& 1.5	& 2	& 3\\
\hline
$R(t)$	& 0.443	& 0.362	& 0.108	& 0.026	& 0.001\\
\hline
$E\;\hat{R}^*(t)$	& 0.349	& 0.278	& 0.076	& 0.018	& 0.001\\
\hline
\%	& 27\%	& 30\%	& 42\%	& 50\%	& 61\%\\
\hline
\end{tabular}
\label{t1_25}
\end{table}

The Resampling method gives an unbiased estimator for this task. The algorithm is obtained for  estimator calculation. Table \ref{t3_25} shows the variance of Resampling estimator 
depending on time $t$.

\begin{table}[ht]
\caption{Variance of the Resampling estimator}
\centering
\begin{tabular}{|l|l|l|l|l|l|l|l|l|l|l|}
\hline
t	& 0.1 & 0.2	& 0.3 & 0.5 & 0.7 & 0.9 & 1 & 1.5 & 2\\
\hline
D R*(t)& 0.079 & 0.081 & 0.085 & 0.098 & 0.108 & 0.110 & 0.108 & 0.091 & 0.082\\
\hline
\end{tabular}
\label{t3_25}
\end{table}

\subsection{Discrete Optimization of Resampling Sample Sizes}

In many practical tasks we need to give recommendations for sample sizes $n_i$. It is clear that we select sample sizes $n_i$ automatically, they must be optimal.

As the Resampling method efficiency criterion is variance, we need to select such $n_i$ values
which minimize variance. Suppose each element of sample $H_i$ has weight $a_i$, $i=1,\ldots,m$,
and the total weight is bound by $b$. Our task is to solve the following optimization task:
\begin{equation}
\mbox{minimize }D(n_1,n_2,\ldots,n_k)
\label{probl}
\end{equation}
by restriction
\begin{equation}
a_1n_1+a_2n_2+\ldots+a_kn_k\le b,
\label{restric}
\end{equation}
where $b$, $\{a_i\}$ un $\{n_i\}$ are integer non-negative numbers, $D(n_1,n_2,\ldots,n_k)$
is the variance of the estimator, which depends on the sample sizes.

In order to solve the given optimization task, we use the dynamic programming method.
Let us consider the function
\begin{equation}
\psi_i(\alpha)=\alpha\sigma_i^2+(1-\alpha)Cov(X_i,X'_i),\qquad i=1,2,\ldots,k,\;0\le\alpha\le 1.
\label{dyn1}
\end{equation}

It can be proved that
\begin{equation}
Cov(\phi_v(X),\phi_v(X'))=\sum_{i\in I_v}\left(\frac{\partial}{\partial x_i}\phi_v(\mu_v)\right)^2\psi_i\left(\frac{1}{n_v}\right).
\label{dyn2}
\end{equation}

It also can be proved that
\begin{equation}
\psi_v(\alpha)=\sum_{i\in I_v}\left(\frac{\partial}{\partial x_i}\phi_v(\mu_v)\right)^2
\psi_i\left(\alpha+\frac{1-\alpha}{n_v}\right).
\label{dyn3}
\end{equation}

We can see that variance $D\;\theta^*$ can be obtained as
\begin{equation}
D\;\theta^*=\psi_k(0).
\end{equation}

Values $\psi_v(\alpha)$ depend on all subnode sample sizes $n_i$.
Let us define these subnodes indexes by $B_v$ and write $\psi_v(\alpha)=\psi_v(\alpha;n_i:i\in B_v)$.

Then the Bellman function, which must be calculated, can be written in the following way:
\begin{equation}
\Phi_v(\alpha,z)=\min_{n_i}\psi_v(\alpha;n_i:i\in B_v),
\label{dyn5}
\end{equation}
where minimization is realized by non-negative integer 
variables $n_i$, which satisfy the restriction
\begin{equation}
\sum_{i\in B_v}a_in_i\le z.
\label{dyn6}
\end{equation}

It can be proved, that the Bellman function can be represented in the following way:
\begin{equation}
\Phi_v(\alpha,z)=\min\sum_{i\in I_v}\left(\frac{\partial}{\partial x_i}\phi_v(\mu_v)\right)^2
\Phi_i\left(\alpha+\frac{1-\alpha}{n_v};z_i\right),
\label{dyn7}
\end{equation}
and minimize it by integer non-negative variables $n_v$ and $\{z_i:i\in I_v\}$,
which satisfy restrictions
\begin{equation}
a_vn_v+\sum_{i\in I_v}z_i\le z.
\label{dyn8}
\end{equation}

At the end the minimal variance $D^*\;\theta^*$ is equal to
\begin{equation}
D^*\;\Theta^*=\Phi_k(0,b).
\label{dyn10}
\end{equation}

In order to calculate the optimal sample sizes $n^*_1,n^*_2,\ldots,n^*_k$, we need to use 
the dynamic programming "forward" procedure.

\underline{Example: Subquery sample size optimization.}
Let we have a query to database which consists of 6 subqueries, as in 
example 1 of section 9.3. 
The execution time of the subquery $i$ is random variable
$X_i$, $i=1,\ldots,6$. The distributions  
$F_i(x)$ of these times are unknown, but only samples $H_i$ are available for each $i$. 

The task is to estimate the expectation of the query working time:
\begin{equation}
\theta = E\;\max(\max(x_1,x_2),\min(x_3,x_4),x_5+x_6).
\end{equation}

Derivatives of all subfunctions $\phi_i$ are calculated. The Bellman functions $\Phi_v(\alpha,z)$ are constructed, formulas (\ref{dyn5}), (\ref{dyn6}), (\ref{dyn7}) and (\ref{dyn8}) are iteratively applied, and formula (\ref{dyn10}) is applied to get an optimal solution.

The obtained results are shown in table \ref{optimiztable}. We can see that the method allows us to decrease the variance of the estimator to 10-40\%.

\begin{table}[h]
\caption{Optimization results}
\label{optimiztable}
\centering
\begin{tabular}{|c|c|c|c|c|}
\hline
$\lambda_i$&$n_i$&$D^*$&$D$&\%\\
\hline
(0.1,0.7,0.2,0.4,0.8,0.5)&(3,3,9,2,2,4,4,9,4,10)&3.37&4.30&22\%\\
\hline
(0.2,0.2,0.4,0.4,0.8,0.8)&(6,6,3,3,3,3,8,4,4,20)&6.03&6.95&13\%\\
\hline
(0.2,0.3,1.0,1.2,0.5,0.3)&(4,4,1,1,9,3,6,1,10,11)&12.59&17.88&30\%\\
\hline
(1.2,0.1,0.3,2.1,0.1,1.5)&(1,1,4,1,12,1,1,4,12,13)&7.64&13.61&44\%\\
\hline
\end{tabular}

\end{table}

\subsection{Point Estimation of Calculation Systems in the Case of Partially Known Distributions}

Suppose the distributions of some variables are known. Variables 
$X_1, X_2, \ldots,X_m$ are given, the distributions of which are unknown (but only 
samples $H_i$ are available), and also variables $Z_1,Z_2,\ldots,Z_\nu$ are given, the distributions of which are known (functions $F_i(x)$ are given).
Function $\phi$ depends on vectors $X$ and $Z$.

The task is to estimate the expectation of function $\phi$, the arguments of which are random variables $X$ and $Z$:
\begin{equation}
\theta = E\;\phi(X,Z).
\end{equation}

The question is the following: how to use the information available from $Z$ knowledge in the most efficient way?

The idea is to use the Hierarchical Resampling method, but make samples from distribution functions. 2 situations are possible:

\begin{itemize}
\item It is possible to calculate the distribution of subfunction $F_{v,x}(y)=P\{\phi_v(x,Z)\le y\}$ in the tree node;
\item It is impossible to calculate the distribution of subfunction in the tree node.
\end{itemize}

\begin{figure}[htbp]
	\centering
		\includegraphics[width=0.75\textwidth]{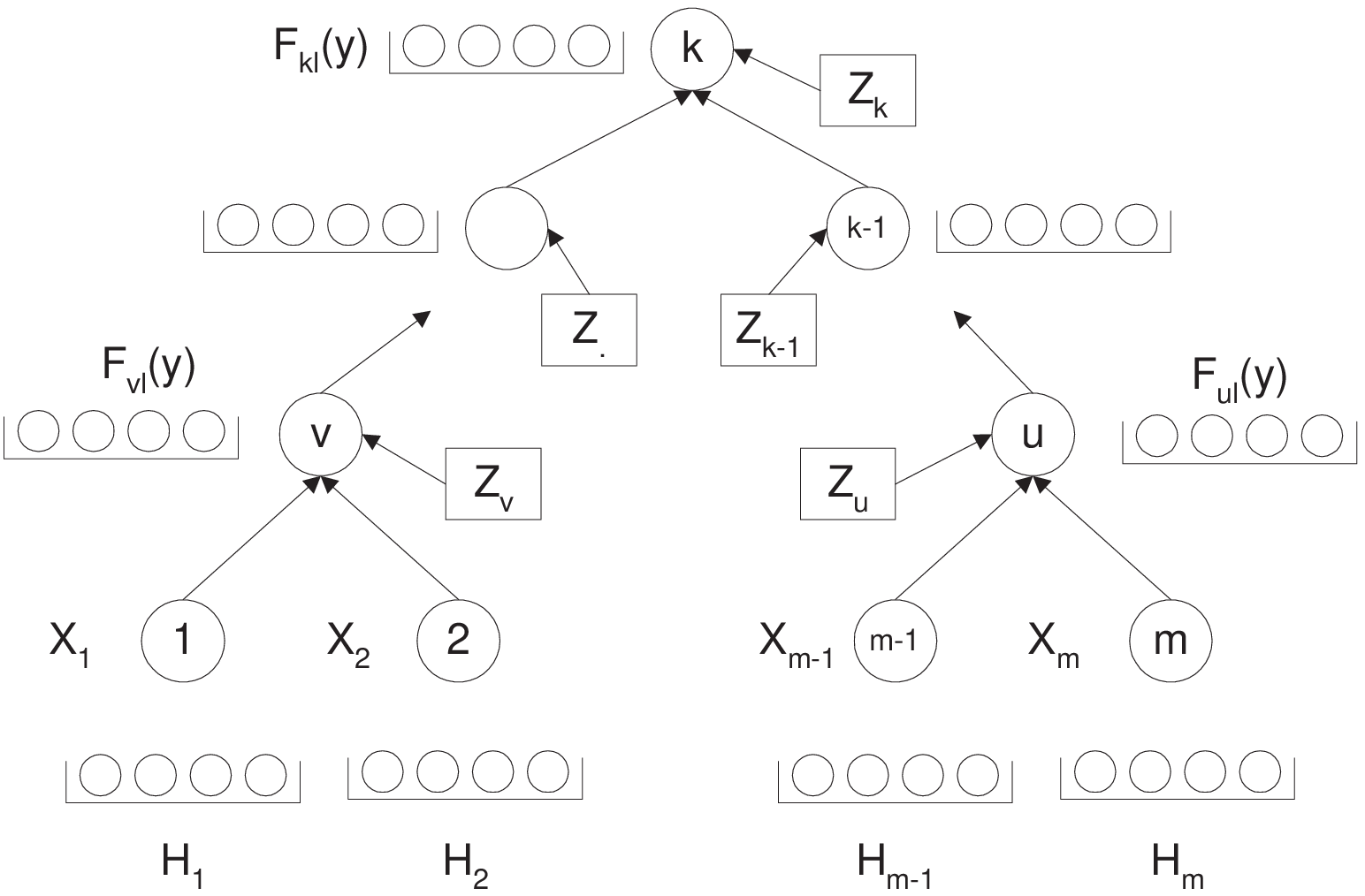}
	\caption{The case of known subfunctions distribution}
	\label{fig:tree55}	
\end{figure}

In the 1-st situation the sample of functions $F_{v,l}(y)$ is constructed in each node, 
where $l$ is a step number, but $x$ is not in index because it is extracted from subsamples.
The calculation tree is shown on the Fig. \ref{fig:tree55}.

At the end the estimator $\theta^*$ is calculated by formula 
\begin{equation}
\theta^* = \frac{1}{r}\sum_{l=1}^r \int\limits_{-\infty}^\infty ydF_{kl}(y).
\end{equation}

\begin{figure}[htbp]
	\centering
		\includegraphics[width=0.75\textwidth]{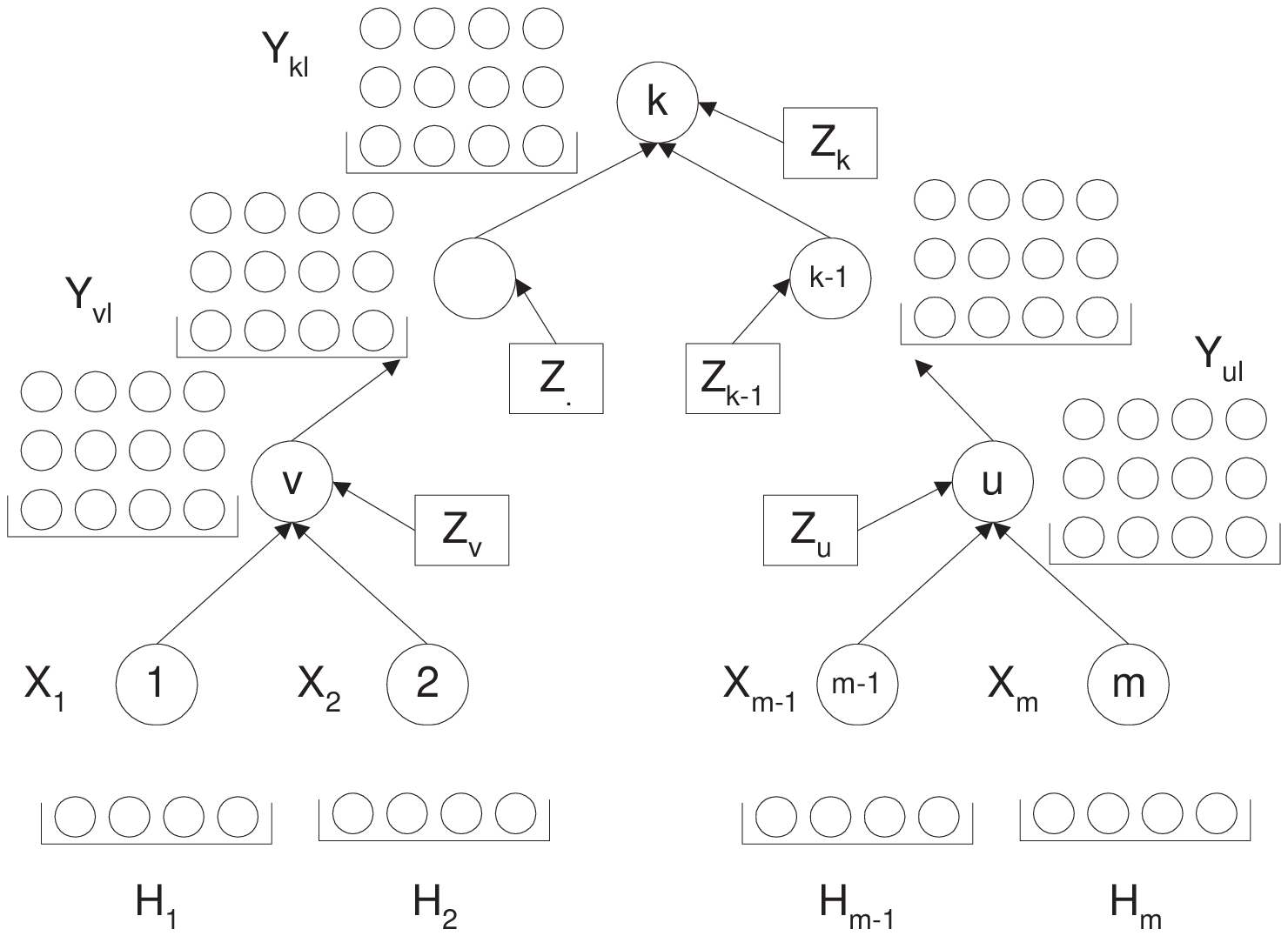}
	\caption{The case of unknown subfunctions distribution}
	\label{fig:tree6}
\end{figure}

In the 2-nd situation we use the $N$-dimensional vector $Y_{vl}$ instead of the function
$F_{vl}(y)$. In order to construct this vector, we select vectors $Y(l,\xi)=(Y_{ij_i(l)\xi})$
from subsamples. Then for each $l=1,2,\ldots,n_v$ and $j$ we construct  
$N$ random variables $\{Z_{jl\xi}:\xi=1,2,\ldots,N\}$. Then we construct
$n_v\cdot N$ vectors $Z(l,\xi)=(Z_{jl\xi}$, calculate values
$$
Y_{vl\xi}=\phi(Y(l,\xi),Z(l,\xi))
$$
and construct vectors $Y_{vl}=(Y_{vl\xi}:\xi=1,2,\ldots,N)$. This procedure is shown on
Fig. \ref{fig:tree6}.

At the end the estimator $\theta^*$ is calculated by formula 
\begin{equation}
\theta^*=\frac{1}{rN}\sum_{l=1}^r\sum_{\xi=1}^N Y_{kl\xi}.
\end{equation}

\underline{Example: Hierarchical query to database.} 
Let us consider the same query to database as in Example 1 of Section 9.3, but with
partially known distributions of subquery working times.
We suppose that the distribution functions $F_2(t)$, $F_4(t)$ and $F_6(t)$ of
2-nd, 4-th and 6-th subqueries working times $X'_2$, $X'_4$ and $X'_6$ are known,
but the distribution functions $F_1(t)$, $F_3(t)$ and $F_5(t)$ of
1-st, 3-d and 5-th subqueries working times $X'_1$, $X'_3$ and $X'_6$ are unknown,
and only samples $H_1$, $H_3$ un $H_5$ are available.
The task is to estimate the probability $R(t)=\theta_t$, that at the time moment $t$
the query will end its work.

In order to follow the above mentioned notation, let us denote $X_1=X'_1$, $X_2=X'_3$, $X_3=X'_5$; $Z_1=X'_2$, $Z_2=X'_4$, $Z_3=X'_6$.
Then our goal is to estimate the expectation of the function $\phi_t$, where $\phi_t$ is the following function:
\begin{equation}
\phi_t(X,Z)=\left\{
\begin{array}{ll}
1 &\mbox{if}\;\min\{\max\{X_1,Z_1\},X_2,Z_2,X_3+Z_3\}>t,\\
0 &\mbox{else}.
\end{array}
\right.
\end{equation}

In this case we have the 1-st situation, when
the conditional expectation $\phi(X_1,X_2,X_3)=\phi(X'_1,X'_3,X'_5)$
is known. It can be calculated as follows:
\begin{equation}
\label{ex3}
\phi(X'_1,X'_3,X'_5)=\left\{
\begin{array}{ll}
0,&\mbox{if}\;X'_3<t,\\
\overline{F_2}(t)\overline{F_4}(t)\overline{F_6}(t-X'_5),&\mbox{if}\;X'_3>t,X'_1<t,X'_5<t,\\
\overline{F_4}(t)\overline{F_6}(t-X'_5),&\mbox{if}\;X'_3>t,X'_1>t,X'_5<t,\\
\overline{F_2}(t)\overline{F_4}(t),&\mbox{if}\;X'_3>t,X'_1<t,X'_5>t,\\
\overline{F_4}(t),&\mbox{if}\;X'_3>t,X'_1>t,X'_5>t.
\end{array}
\right.
\end{equation}

Formulas for $\mu_{11}$ value calculation are obtained. Variance dependence on different parameters is analyzed, different cases are compared and it is shown that the method is
relatively effective for solving this task. The comparison of the Resampling with unknown and partially known distributions is shown on Fig. \ref{draw5_4n}.

\fig{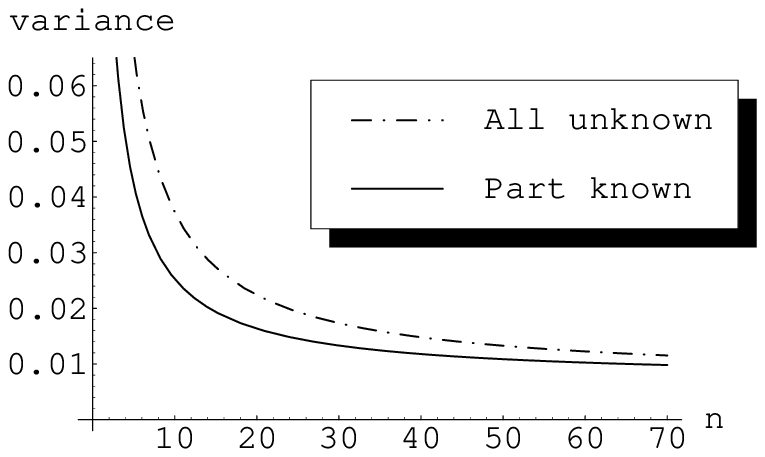}{Comparison of Hierarchical Resampling method for unknown and partially known distributions}{draw5_4n}

\subsection{Resampling Interval Estimation of Logical Systems}

Before we considered the point Resampling estimators.
But in many practical tasks it is important to know the interval, where the value of parameter traps with the given probability. In this case we have to deal with interval estimation.

Let we have a function $\phi(X_1,X_2,\ldots,X_m)$ of $m$ random variables. 
The task is to construct the confidence interval with level $\gamma$ for the function $\phi(X_1,X_2,\ldots,X_m)$ expectation $\theta=E\;\phi(X_1,X_2,\ldots,X_m)$.

Using the Resampling method, we can estimate the expectation $\theta^*$ of the function $\phi$.
We make $r$ such realizations $(\theta^*_1,\theta^*_2,\ldots, \theta^*_r)$.
We order this sequence, obtaining order statistics $\theta^*_{(1)},\theta^*_{(2)},\ldots, \theta_{(r)}$.
Accept $(\theta^*_{(\lfloor\alpha r\rfloor)},\infty)$ as $1-\alpha$ upper confidence interval for the parameter $\theta$. Here $\lfloor\alpha r\rfloor$ means a greater integer number which is less than or equal to $\alpha r$.

Due to the vector $(\theta^*_{(1)},\theta^*_{(2)},\ldots, \theta_{(r)})$ component dependence the coverage probability of the parameter $\theta$ by the interval $(\theta^*_{(\lfloor\alpha r\rfloor)},\infty)$ differs from $1-\alpha$.
The task is to calculate the actual coverage probability 
\begin{equation}
R = P\{\theta^*_{\lfloor(\alpha r)\rfloor}\le \theta\}.
\end{equation}

The method is described in a paper of Andronov, Fioshin \cite{AndrF_04}.

Suppose the function $\phi(X_1,X_2,\ldots,X_m)$ depends on the order of $X_i$ only, not on the actual values. The idea is to fix this order and to find the conditional probability $R = P\{\theta^*_{\lfloor(\alpha r)\rfloor}\le \theta\}$ on the condition, that order is given.
The disadvantage of this approach is the large dimension of the task, because the number of different combinations can be large. In order to decrease the dimensions, a protocol definition is proposed.

At the beginning let us describe the protocol definition in the case of 2 dimensions.
Suppose the function $\phi(x_1,x_2)$ depends on 2 arguments.
We have 2 samples $H_1 = (X_{11},X_{12},\ldots,X_{1n_1})$ and
$H_2 = (X_{21},X_{22},\ldots,X_{2n_2})$.
Let us order both samples and calculate, how many elements of the second sample are between the first sample neighbor elements:
\begin{equation}
c_i = \#\{X_{2j}:X_{1(i)}<X_{2j}\le X_{1(i+1)}\},
\end{equation}
where $\#X$ means the number of elements in set $X$.

We can find the probability of each such protocol.
If we know the protocol, we can calculate the conditional coverage probability and then the coverage probability $R$.

In the multidimensional case the protocol is defined in analogous way.
Suppose the function $\phi(x_1,x_2,\ldots,x_m)$ depends on $m$ arguments;
we have $m$ samples $H_i$.
We order elements of all samples and write a number of a sample, which the each element belongs to:
\begin{equation}
c_j = i \Leftrightarrow X_{(j)} \in H_i.
\end{equation}

For example, if $H_1=(2.5,6.3,1)$, $H_2=(0.5,4.7)$, $H_3=(3.1,0.2,5.2)$, then the ordered sequence is $(0.2, 0.5, 1, 2.5, 3.1, 4.7, 5.2, 6.3)$ un $c=(3,2,1,1,3,2,3,1)$.
We can calculate the probability of each such protocol, the conditional coverage probability and the coverage probability $R$.

\underline{Example 1: Minimal-time process selection.}
Let us have an information system which controls processes.
It is known that the optimal strategy of such system is to execute the shortest processes first. 

Let we have $m$ processes in the system. 
We suppose that the processes execution times are independent random variables
$X_1,X_2,\ldots,X_m$. The distributions $F_1(x),F_2(x),\ldots,F_m(x)$ are
unknown, but only sample populations $H_1,H_2,\ldots,H_m$ are available for
each $X_i$.

We suppose that the system selects the process the execution time of which is predicted to be minimal;
the system gives a number $m$ to this process.
This means that the system supposes that $X_m<\min(X_1,X_2,\ldots,X_m)$. Our task
is to estimate the probability of the correct selection:
\begin{equation}
\label{ex61_1}
\theta=P\{X_m<\min(X_1,X_2,\ldots,X_m)\}.
\end{equation}

We also need to construct the upper confidence interval for $\theta$ with a given 
confidence level $\gamma$.

The corresponding protocols were constructed. The probability of each protocol was calculated, the conditional coverage probability found. It allowed us to find an actual coverage probability $R$. The results of the calculation are presented in Table \ref{t6_1}.

\begin{table}[ht]
\caption{Actual coverage probabilities}
\label{t6_1}
\centering
\begin{tabular}{|c|c|c|c|c|c|}
\hline
&\multicolumn{5}{c|}{Coverage probability $R$}\\
\hline
$(n_1,n_2,n_3)$	&$\gamma$=0.5	&$\gamma$=0.6    &$\gamma$=0.7	&$\gamma$=0.8	&$\gamma$=0.9\\
\hline
(3,3,3)		&0.533	&0.576	&0.625	&0.686	&0.770\\
(9,9,3)		&0.519	&0.571	&0.630	&0.701	&0.793\\
(4,4,4)		&0.521	&0.578	&0.640	&0.709	&0.797\\
(6,6,4)		&0.516	&0.576	&0.642	&0.715	&0.807\\
(5,5,5)		&0.515	&0.579	&0.646	&0.722	&0.817\\
(3,3,8)		&0.516	&0.581	&0.651	&0.728	&0.823\\
(4,4,7)		&0.512	&0.580	&0.652	&0.732	&0.830\\
\hline
\end{tabular}
\end{table}

\underline{Example 2: Process ordering.}
Suppose like in the previous example we have an information system which controls processes.
The system orders the processes by the estimated execution time.
The system gives corresponding numbers to ordered processes: this means that the system supposes that $X_1<X_2<\ldots<X_m$. The goal is to estimate the probability of the correct ordering
\begin{equation}
\label{ex62_1}
\theta=P\{X_1<X_2<\ldots<X_m\}.
\end{equation}

We also need to construct the upper confidence interval for $\theta$ with a given 
confidence level $\gamma$.

The corresponding protocols were constructed. The probability of each protocol was calculated, the conditional coverage probability found. It allowed us to find an actual coverage probability $R$. The results of the calculation are presented in Table \ref{t6_2}.

\begin{table}[ht]
\caption{Actual coverage probabilities}
\label{t6_2}
\centering
\begin{tabular}{|c|c|c|c|c|c|}
\hline
&\multicolumn{5}{c|}{Coverage probability $R$}\\
\hline
$(n_1,n_2,n_3)$	&$\gamma$=0.5	&$\gamma$=0.6    &$\gamma$=0.7	&$\gamma$=0.8	&$\gamma$=0.9\\
\hline
(3,3,3)		&0.593	&0.635	&0.680	&0.730	&0.803\\
(9,9,3)		&0.524	&0.595	&0.675	&0.762	&0.862\\
(4,4,4)		&0.540	&0.606	&0.677	&0.757	&0.848\\
(6,6,4)		&0.525	&0.600	&0.678	&0.766	&0.864\\
(5,5,5)		&0.523	&0.601	&0.682	&0.770	&0.866\\
(3,3,8)		&0.536	&0.604	&0.678	&0.760	&0.855\\
(4,4,7)		&0.522	&0.600	&0.681	&0.769	&0.866\\
\hline
\end{tabular}
\end{table}

\section*{Conclusions}
\addcontentsline{toc}{section}{Conclusions}

In the present work the properties of the Resampling method were analyzed and the possibility
of its application to the information systems estimation was studied. Different Resampling
method application cases and tasks were analyzed, such as simple Resampling, Hierarchical
Resampling, Resampling in the case of partially known distributions, sample size optimization, confidence interval construction.

For each of the mentioned situations or tasks the methodology and algorithms of the Resampling method application were shown. It was shown how to calculate the values of the method efficiency criteria.

For each of the mentioned tasks or situations examples from the information systems area were analyzed, and the Resampling method was applied for the systems estimation. For each class of the task the methodology of the Resampling method application was shown, algorithms were obtained for the method efficiency calculation, a number of examples illustrate the dependence of different factors on the efficiency of the method, and a comparison of various methods was made.

From the obtained results it is possible to conclude that the Resampling method can be a good alternative to the classical methods in the case of information systems analysis.

The methodology that is obtained in the present work and other results can be a basis of the
software that performs system simulation and estimation using the Resampling approach.

\renewcommand{\refname}{Publications with Author Participation}

\end{document}